# OPERATIONS MANAGEMENT OF SATELLITE LAUNCH CENTERS[1]

Andrea Tortorelli
Sapienza University of Rome, Via Ariosto 25, 00185, Rome, Italy, +39 0677274037,
tortorelli@diag.uniroma1.it

Alessandro Giuseppi
Sapienza University of Rome, Via Ariosto 25, 00185, Rome, Italy, +39 0677274037,
giuseppi@diag.uniroma1.it

Federico Lisi
Sapienza University of Rome, Via Ariosto 25, 00185, Rome, Italy, +39 0677274037,
lisi@diag.uniroma1.it

Emanuele De Santis
Sapienza University of Rome, Via Ariosto 25, 00185, Rome, Italy, +39 0677274037,
desantis.1664777@studenti.uniroma1.it

Francesco Liberati
Sapienza University of Rome, Via Ariosto 25, 00185, Rome, Italy, +39 0677274037
liberati@diag.uniroma1.it

**Abstract**

Driven by the business potentialities of the satellite industry, the last years witnessed a massive increase of attention in the space industry. This sector has been always considered critical by national entities and international organizations worldwide due to economic, cultural, scientific, military and civil implications. The need of cutting down satellite launch costs has become even more impellent due to the competition generated by the entrance in the sector of new players, including commercial organizations. Indeed, the high demand of satellite services requires affordable and flexible launch. In this context, a fundamental aspect is represented by the optimization of launch centers' logistics. The aim of this paper is to investigate and review the benefits and potential impact that consolidated operations research and management strategies, coupled with emerging paradigms in machine learning and control can have in the satellite industry, surveying techniques which could be adopted in advanced operations management of satellite launch centers.

## 1. Introduction

In the last two decades, the space race has witnessed a significant and renewed growth in interest due military, civil and economic interests. The main driver of such growth of interest is represented by the business potentialities offered by the satellite industry [1], [2], [3]. Indeed, besides governmental bodies, many new commercial and semi-commercial entities entered the market (e.g. Space X, Blue Origin, RocketLab and Virgin Orbit), lured by the new business opportunities. Much attention has also been devoted to the Launch Manufacturing and Services segments [4], which are the enabler for the satellite industry: in the next decade, these segments will account for an estimated market of $255bn [3].

Affordable and flexible launch services (i.e. satellite manufacturing and launch logistics) are needed, increasing launch rates and decreasing the deployment times [1], while keeping high levels of quality and safety. For these purposes, the following aspects are crucial: (i) the launcher production, storage and transport and (ii) the logistics of the launch center. Following on these considerations, this paper will investigate how to integrate in the launch centers logistic Predictive Analytics techniques, advanced Business Intelligence (BI) and Operations Management (OM) methodologies for the optimization of launch centers operations. The paper will present an introduction to the main definitions and a high-level review of several techniques appeared in literature in advanced operations management and supply chain optimization, which could be of interest in the context of improving the logistic of satellite launch centers. One of the key innovation challenges for example is in understanding how the most recent developments enabled by big data analytics, machine learning and advanced control techniques can be best used for the optimization of satellite launch centers (e.g. by integrating predictive analytics with real time optimization of the operations). The remainder of the paper is structured as follows: Chapter 2 presents a brief description of launch centers, highlighting the current barriers, trends and opportunities as well as the notion of launch centers' logistic management; Chapter 3 reviews algorithms, techniques and methodologies which are, or could be, adopted for the optimization of launch centers operations; Chapter 4 provides an insight on how the reviewed solutions can be embedded in launch centers operations management; finally, Chapter 5 concludes the work.

[1] This work has been carried out in the framework of the SESAME project, which has received funding from the European Union's Horizon 2020 research and innovation programme under grant agreement No 821875. This work reflects only the author's view. The EU Commission and the Research Executive Agency are not responsible for any use that may be made of the information it contains.

## 2. Launch centers

Launch centers are one of the key enablers of the Satellite industry and represent a fundamental part of the satellite supply chain. Since the first satellite launch in 1957, only 27 launch centers have been used (of which 5 are not active anymore) [5]. It is clear that the optimization of the launch centers logistics plays a crucial role in containing costs and increasing the flexibility of launch operations, which is key in order to match the actual and forecasted growing demand for satellite services.

The focus of this paper is on the optimization of launch centers operations. The term launch center covers the launch vehicles (i.e. all its components, including the payload) and the ground infrastructure. The choice of not using the term spaceport has been made to stress the area of interest of the present work. Operations cover all those resources (including personnel, time and facilities) and activities (including assembly, monitoring and maintenance) required to put in orbit launch vehicles. This definition of operations is in line with the one adopted in [6]. More specifically, launch centers operations cover the following phases: the arrival of the launch vehicle components, the assembly, checkout, storage and integration phases, the launch and post-launch operations. Figure 1 shows the France and European Kourou Guaiana launch site details. As depicted in the figure, launch centers are complex systems envisaging internal transportation infrastructures, warehouses, assembly and integration buildings and control rooms.

With the growth of the so-called New Space (or Space 2.0), the satellite industry experienced significant changes with a direct impact on the nature and role of launch centers [7]. On the one hand, the high demand of public and private organizations highlighted the business potentialities of launch centers. These potentialities captured the interest of both private and national institutions due to the economic return and social, cultural and scientific positive repercussions on the territory [8], [9], [10]. Indeed, in 2017 commercial spacecraft deployed witnessed an increase of 200% [11] whereas the global space economy saw an average annual increase of 6,7% from 2005 to 2017 (considering both public and private budgets) [12]. The optimization of launch centers operations can give a fundamental contribution to reducing the cost of access to space, which is the most relevant barrier in the space industry [13]. In this perspective, the ability of operations cost prediction has been identified as crucial and many attempts at cost modelling and cost metrics definition have been presented [6], [14], [15]. The next chapter presents basics notions about launch centers operations.

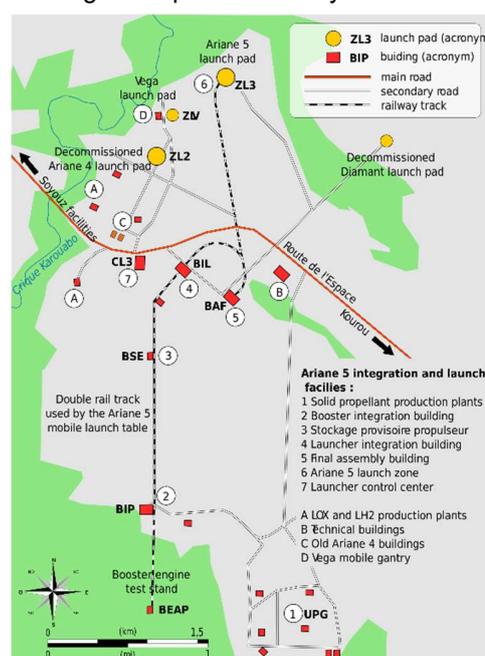

### 2.1. Logistic of Launch Centers

There are several definitions in literature of launch system operations. In [6], "operations" is defined as "the facilities, ground support equipment, people, and time required to perform all activities necessary to maintain the launch system on the ground and in space". The authors present an overview of NASA's lunch systems operations in the 2000s (which is still actual in some aspects today), and discusses goals and challenges of the transition towards NASA's 2nd and 3rd generation of operations models. Traditional operations models are characterised by a combination of planned (e.g. removal and replacement of limited life components, refuelling, testing and checkout of systems, etc.) and unplanned (related to anomalies, failures, etc.) activities, which are decided and scheduled during regular or ad hoc planning meetings. There is a lack of comprehensive information system collecting, correlating and exploiting data from the launch system. The planning, operation and maintenance methods and technics used are basic and conservative, mostly paper-based and static, i.e. unable to adapt to the changing status of the launch systems, and therefore necessarily conservative, to ensure safety. This results in the over dimensioning of the system and underutilization of resources. The second and third generation of operations models aim at improving launch center performances (reduce costs, reduce maintenance time, increase number of missions) through the introduction of more advanced diagnostic and prognostic techniques (e.g. leading to less conservative and real time life/health estimation techniques), adaptive logistic and planning models which provide real time decision support to the operations players, automated determination of maintenance corrective actions to be put in place. This is made possible first of all through the extensive adoption of paper less and real time information systems able to gather data from the launch system, store it (including the concept of digital twin of each component) and aggregate, process and correlate it to derive actionable analytics for the operations.

*Figure 1: Kourou Launch Site details*

To propose and implement advanced operations models it is necessary to develop models able to capture the complexity of the launch systems at the level of detail needed. Reference [16] presents relevant ideas to develop models able to capture and model the overall availability of the various launch subsystems, including the reliability (the probability that an item will work with no failures during a given period of time) and the maintainability ("the probability of completing a repair within a specified time" [16]) functions. In [16], the authors also stress the important aspect, also to be taken into account in advanced logistic algorithms, that reconfigurations and improvements, modifications in the launch system design require also to check again the consistency of the availability requirements assigned to the different subsystem.

In [17], an interesting overview of EU's VEGA program at the Guyana Space Center is presented and the key systems, facilities and procedures involved in the VEGA operations are described. An interesting aspect to be considered in this case is the need to plan the operations taking into account that parts of the resources are shared across multiple launch programs running in parallel in the same spaceport (VEGA, ARIANE, SOYUZ). The paper also presents the key improvements planned in the upgraded launch systems VEGA-C, many of which involve refinements of the operations, with the goal of increasing throughput and reduce mission time. The proposed improvements include: (i) performing part of the integration, finalization and acceptance of specific components outside of the perimeter of the launch campaign, e.g. directly at the assembly lines (in Europe) or in other facilities of the space port, so as to decouple tasks and put them as much as possible in parallel, to save time, (ii) increase in parallelization of tasks, (iii) increase of automation procedures, (iv) reorganization of the checks and reduction of duplicated/overlapping checks and tests. The target is to reduce the overall launch campaign duration to 20 working days.

Similarly, [18] presents an overview of the next generation ARIANE 6 launch system, presenting interesting aspects of the launch system design and operation, discussing also the improvements beyond the current ARIANE 5 generation put in place to reduce costs, increase versatility and flexibility of operations (supporting both institutional and commercial customers) and achieve a target of 12 launches per year. The paper also discussed the adopted methodology for systems analysis, optimization and control and the ARIANE 6 high level operational concepts. In addition to the issues already mentioned, the paper also stresses the importance of carefully designing the requirements verification and validation phases (to be done at the lowest possible level, to early detect issues) and the careful design of the assembly phases, in order to avoid as much as possible that one phase constraints the other (which can be inefficient and particularly problematic in case of anomalies and the consequent need for de-assembling for fixing/replacing parts).

## 3. Review of Launch Centers Operations

Several papers are available in literature reviewing relevant aspects of logistics and operations management which can be of intereste for advanced operations of launch centers. [19] and [20] are fundamental contributions presenting a review of modelling and control/optimization techniques for the supply chain management problem. Reference [21] presents a review of the operational control techniques for discrete event logistics systems (DELS), modelling and tackling DELS operations as a combination of the so-called admission, sequencing, assignment, changing-state tasks. In the following, we discuss a selection of technical works presenting different approached to operations optimization.

### 3.1. Model Predictive Control Techniques

Reference [22] applies model predictive control (MPC) to the optimization of the operations of a supply chain network (customers, retailers, distributors, warehouses, production facilities, and suppliers), modelled using the Mixed Logic Dynamical (MLD) methodology, which integrates dynamical models (e.g. to model actual systems/phenomena, like the piling up of inventory) and logic constraints (to capture inherent logic constraints, operational modes, etc.). Customers' demand is included as a disturbance and the maximization of the overall profit is sought via proposed centralised, decentralised and semi-decentralised models. This approach is potentially very interesting for the launch systems optimization, as launch systems can be seen indeed as supply chains. Reference [23] applies MPC to the optimization of semiconductor manufacturing. MPC is here applied to ensure that the targets decided at strategic and tactic levels (e.g. inventory targets) are optimally tracked and ensured by the operations. An interesting case study analysed is the optimization of the operations of multi-product manufacturing, which is of interest for the launch systems operating multiple launch vectors.

Several other references investigate MPC in the context of operations management and inventory control. Among the recent ones, we mention [24], which explicitly introduces economic MPC, a variant of MPC in which the MPC objective function directly includes economic performance indicators, so that the control will result in the optimization of the economic performance of the controlled system. For example, the method in [24] seeks to minimize the expenses associated with inventory operations management in the pharmacy department of an hospital.

### 3.2. Closed-loop logistics models for supply chain management

With the increasing attention to environmental issues, closed-loop logistics has gained a significant attention in the last years [25]. The methodologies adopted in this context, aimed at minimizing the resource usage while maximizing efficiency, provide useful inputs for the launch centers operations management. Indeed, launch centres operations include also post-launch operations and, furthermore, due the variety and complexity of such operations, the presence of feedback/feedforward loops will certainly provide benefits. In particular, the concept of Reverse Logistics, together with predictive maintenance and quality, appears to be particularly suited for the launch centres context in which high-efficiency, rescheduling abilities and flexibility are fundamental aspects.

In [26], the authors developed a model for solving the cyclic logistics network problem and adopted a spanning tree based genetic algorithm. In [27], the authors developed a quantitative modelling for decision support addressing modularity, reparability and design of the logistics network. An interesting work in this context is [28], where the authors addressed the closed-loop logistics problem in the context of hybrid manufacturing and remanufacturing facilities under finite-capacity hybrid distribution and collection centers. In [29] an exhaustive review of this class of methodologies is presented.

### 3.3. Cooperative Control and Distributed Approaches

Supply chains and logistics are more and more growing in complexity, so that the classical centralised ICT and control tools may lack the flexibility and the scalability needed. For this reason, researchers and practitioners are increasingly exploring new approaches based on cooperative control. Reference [30] provides an excellent overview of the key motivations and concepts behind cooperative control in logistics. Several possible control approaches are presented, such as multi agent control, game theory, learning-based approaches and distributed MPC. As an example, [31] presents a compressive view of the problem Prognostics and Health Management (PHM) for large and geographically distributed supply chains. The authors show how recent learning techniques can be integrated in the operations to achieve optimal Prognostics and Health Management with tight integration in the operations optimization routines, which includes the adoption of distributed optimization techniques in the large scenarios.

### 3.4. Data Driven Methods

The amount, complexity and heterogeneity of data and interactions, together with the high-quality, high-efficiency and zero-errors requirements characterizing the satellite industry, highlights the utility of adopting Big Data Analytics for launch centers operations. Such techniques have already been applied in several sectors and their benefits for supply chain management have already been proven. In [32], for example, the authors discuss the potential impact of the adoption of Big Data Analytics techniques for operations management in industrial environments. More specifically, the authors discuss how the integration of soft sensors in such domain will lead to an increase of reliability by means of predictive analytics techniques. This aspect has been also discussed in [33], where the authors proposed a definition of logistics predictive analytics and discussed their impact on business operations. In [34], the authors focus on Big Data Business Analytics and discuss its application to logistics and supply chain management. Furthermore, the authors propose a classification of such techniques based on the nature of the analytics and identify three main categories: predictive, descriptive and prescriptive. An interesting discussion on the integration of these methodologies in a holistic supply chain management is also presented.

In [35] it is discussed the impact of the integration of cyber and physical devices in industrial environments and the difficulties of managing such hybrid systems. The authors underline the benefits of the adoption of Big Data Analytics techniques and also propose a generic architecture for decision support able to collect and process industrial data in a centralized and automated way. Similarly, in [36] the authors discusses on how to fill the gap between data science and supply chain management and propose a framework for integrating big data analytics and supply chains. [37] presents an extensive analysis of Big data analytics techniques in the context of supply chain management proposing a new rationale for classifying them. Based on such classification and a review of several application areas, the authors point out how in closed-loop logistics (see Section 3.2) big data analytics techniques are scarcely used. Furthermore, it is pointed out that, in the context of supply chain management, more attention has been devoted to prescriptive analytics than predictive analytics. In [38], the authors underline the relevance of the latter aspect in the context of Industry 4.0 and more specifically discuss the benefits of predictive maintenance for increasing efficiency and providing reliable and high-quality services. Similarly, in [39] the authors present a data-driven approach for solving a procurement and logistics problem in a sustainable way. It is showed how the integration of big data in the supply chain modelling allows to capture and deal with real time changes in the environment. It is also discussed how such integration increases the computational costs of the model and therefore a heuristic is proposed for finding sub-optimal solutions.

### 3.5. Digital Twin

Key for the implementation of optimized logistic and operations methodologies is the derivation of models and representation of the systems and processes to be optimised. A relevant trend observed in this sense, also in the context of the new Industrial 4.0 paradigm, it that of the "Digital Twin". By digital twin it is meant a virtual replica of the system under control (single components, systems or complete processes), which captures all the relevant aspects of the system (e.g. going from the geometrical shape of a manufacturing tool to the mathematical models or rules and procedures associated to or relevant for the replicated component) and is dynamically kept aligned with the evolution of the actual system. Sets of digital twins can be thus organised into a virtual space which represents to the operator an accurate and easily accessible replica of reality, which can be used first of all to have always awareness of the status of the system, and also notably to perform studies and optimization procedures first on the virtual system, and then on the real one, after a decision making/selection process. Digital twins are coupled to the real system via a sensor and information layer/cloud which keeps the status of the real and the virtual world aligned. [40] presents a comprehensive discussion of the modelling and operations of digital twins in the context of manufacturing, clearly explaining the potentiality of the approach, and presenting as well a real case study from the aerospace manufacturing industry, showing how the use of digital twins can e.g. reduce defects and reworking (by performing quality tests on digital twins during production, and triggering proactive condition-based corrective actions on the real parts being manufactures) and production times and inventory accumulation (by optimizing logistics and operations in the digital replica of the system first and then implementing them on the real system). It is worth stressing that advanced solutions will propose a close interplay between the real system/supply chain and their digital representation of it, and the interaction is expected to be sequential (i.e. test and apply pattern) only in simple real cases. Instead, the digital twin will be aligned in real time to the real status of the system and automatic control/logistic procedures will: i) plan the optimal allocation and control of resources and ii) react in real time to unforeseen events (e.g. deriving from faults, uncertain inputs, etc.), in an optimal way that is pre-tested, or tested in near real time on the digital twin system. This is expected to improve the flexibility of the system and reduce operational and capital costs, as a result of the more efficient use of resources. In [41] it is discussed the use of digital twin technology for Digital smart production management and control of complex product assembly shop-floor. The paper provides also an interesting discussion of the evolution of shop floor production management and control strategies from traditions (passive, reactive) approach to real-time (real time data acquisition and control technologies implemented), predictive, and finally proactive, in which the system can not only infer the real and near future status of components, services, etc. (i.e. the predictive stage) but is also intelligently uses this information to best adjust, react, and in general control the future decisions and evolutions of the process. The paper provides a real case study of the application of the digital twin concept to the modelling and optimization of a satellite assembly shop-floor. Concluding, the concept of digital twin is meeting huge interest in research and industries, with many works (see [42]) showing how it can effectively augment the physical system into a cyber-physical system space in which advanced digital and control routines can be deployed to improve quality, cost performance, reliability and other key performance indicators for the system.

## 4. Optimization of Launch Centers operations

In the previous chapter it has been performed a state-of-the-art review of methodologies, models and techniques adopted in the context of logistics and supply chain management. In Chapter 2 it has been discussed the relevance and potential impact on the satellite industry of optimizing launch centres operations. The present chapter is focused on showing of how the reviewed solutions can be embedded in the reference context for the optimization of launch centers operations. Figure 2 shows a schematization of the activities involved in launch centers and where the reviewed algorithms can be embedded.

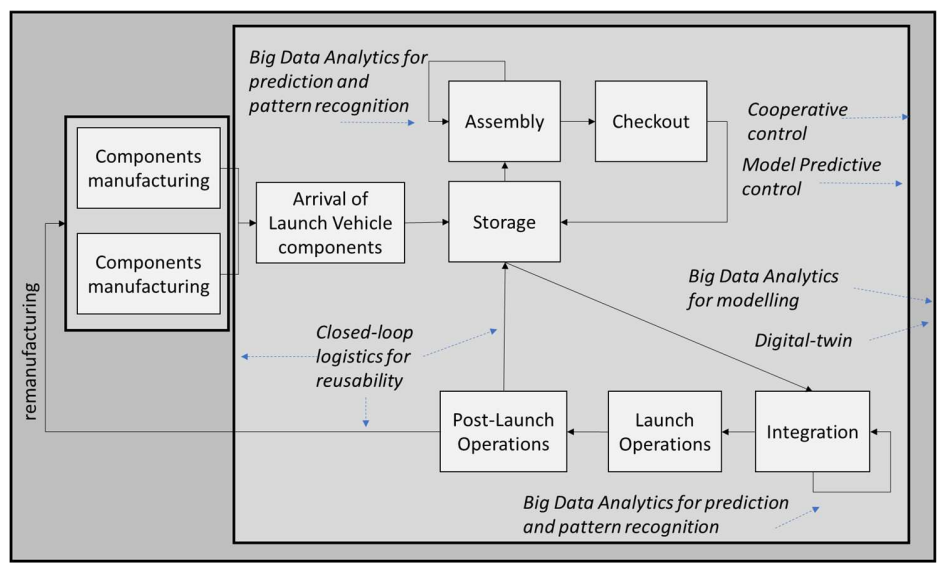

*Figure 2: Schematization of Launch centres activities*

As it can be observed, Model Predictive (Section 3.1), Cooperative, Distributed (Section 3.3) control approaches can be useful for the optimization of launch centers operations. More in detail, Cooperative and distributed control techniques can be applied also in a more general context for the optimization of the whole launch vehicle supply chain, from the components manufacturing to the post-launch operations. Closed-loop logistics models (Section 3.2) can be successfully applied for implementing a comprehensive operations management considering also the notion of reusability which has been identified as one of the most promising solutions for launch costs reduction. Data-driven techniques (Section 3.4) can be used for increasing the performances of several activities. First, the integration of Data Analytics techniques in the modelling process will allow to implement a reactive operations management tool able to address real-time changes in the environment. Second, the adoption of such techniques for support during the assembly and integration phases will allow to induce pattern recognition and predictive capabilities in the management process. More specifically, it would be possible to embed predictive maintenance and quality algorithms in the decision process thus fostering the reduction of deployment times. Finally, the concept of digital-twin (Section 3.5) represent a powerful tool for situational awareness and simulations.

## 5. Conclusions

In the present paper it has been addressed the problem of launch centers operations optimization. It has been discussed crucial role of launch centers in the context of the satellite industry and how, given the current trends in satellite services demand, they could represent a bottleneck for the explosion of such sector. Driven by these considerations, a literature review has been performed with the aim of identifying the most promising and suitable methodologies that, when embedded in the launch centers operations management, enable an increase of launch services flexibility and affordability (which have been identified as the main barriers in the current and future satellite industry). The integration of the reviewed solutions in launch centers operations management and their potential impact has also been discussed. The Space industry is experiencing significant changes and it represents a crucial sector for governmental entities as well as for commercial organizations. Being able to integrate advanced solutions in the satellite industry supply chain management will allow to fully exploit the potentialities of such sector. Topics that should be further analysed include the launch centers supply chain in its entireness (including components manufacturing and regulations), the development of models for launch centers logistics enabling predictive and adaptive management capabilities and the actual implementation in such context of the solutions discussed.

**Acknowledgements**

The authors gratefully acknowledge A. Pietrabissa and S. Mascolo from CRAT, and the members of the SESAME consortium, for the fruitful discussions and contributions.